\documentclass[12pt]{article}
\usepackage{graphics,epsfig}

\newcounter{saveeqn}

\begin{document}

\begin{titlepage}\hfill HD-THEP-09-04\\[5 ex]

\begin{center}{\Large \bf Lowest order covariant  averaging of a 
perturbed  
metric and of the Einstein tensor} \\[5 ex]

{\bf Dieter Gromes}\\[3 ex]Institut f\"ur
Theoretische Physik der Universit\"at Heidelberg\\ Philosophenweg 16,
D-69120 Heidelberg \\ E - mail: d.gromes@thphys.uni-heidelberg.de \\
 \end{center} \vspace{2cm}

{\bf Abstract:} We present an explicit averaging formula in lowest order.
Besides an arbitrary smearing function it contains two integrals of this 
function. This is necessary in order to achieve covariance. There is no 
need to solve any equations. In three dimensions the same averaging 
formula yields a covariant averaging of the Einstein tensor and thus of 
the field equations. We also present a  simple extension to static 
perturbations in four dimensions. Various further extensions of the formalism 
appear possible.

 \vfill \centerline{April 2009}

\end{titlepage}

\setcounter{equation}{0}\addtocounter{saveeqn}{1}%

\section{Introduction}

The averaging problem  in general relativity was first raised by 
Shirkov and Fisher \cite{ShirFi} in 1963. The energy-momentum-tensor used in
cosmological models is an average over the non homogeneous  
tensor present in nature. But, due to the non linear nature of Einstein's 
equations, the metric  belonging to the averaged energy-momentum tensor is not 
identical to the averaged metric. The averaging prescription by itself 
provides a fundamental problem because of the freedom of choice of 
coordinates. The authors of \cite{ShirFi} suggested to
integrate the metric tensor over a four dimensional volume with the 
familiar factor $\sqrt{-g}$ in the measure. Such an expression is, 
however, not 
covariant due to the freedom of performing local transformations.
A covariant averaging prescription can be constructed by introducing a  
bivector $g_\alpha^{\beta }(x,x')$ of 
geodesic parallel displacement, as discussed 
in the appendix of \cite{Isaac}. This
transforms as a vector with respect to coordinate transformations at
either $x$ or $x'$ and maps a vector $A_\beta (x')$ to 
$\bar{A}_\alpha(x)= g_\alpha^{\beta }(x,x') A_\beta(x')$, analogously for 
higher order tensors.
An  averaging with the help of bivectors was also
used in the work of Zalaletdinov \cite{Zalat1} where the emphasis was on the 
commutativity of averaging and covariant differentiation. As remarked by 
Stoeger, Helmi, and 
Torres \cite{StoeHT} the method of using a covariantly conserved 
bivector is not applicable to the metric, because 
the covariant derivative of the metric vanishes. The metric is therefore 
invariant under this averaging procedure.
In the  thesis of Behrend \cite{Behrend} the metric is represented by 
tetrads and the averaging performed over the latter. The tedrads are chosen 
according to a covariant minimalization prescription. 

This is only an extremely brief survey of the literature. For more references
as well as the implications for the fitting problem, back reaction, 
contributions to dark energy, we  refer e.g. to the monograph of 
Krasinski \cite{Kras} and
to the comprehensive recent review of Buchert \cite{Buchert}.

 Under a covariant averaging process 
we understand a prescription which has the following properties. 
Let two observers describe the same physics in different 
coordinate systems $S$ and $S'$, with metric tensors 
$g_{\mu\nu}$ and $g'_{\mu\nu}$. Both 
of them apply a definite averaging procedure in their respective systems,
resulting in the averaged metrics\newline  
$<g_{\mu\nu}>$ and $<g'_{\mu\nu}>$, respectively.
Then the results have to be connected by the same transformation as the 
original metric, i. e.

\begin{equation} 
<g'_{\mu\nu}> = <g_{\mu\nu}>'.
\end{equation}
In other words, the operations of averaging and of coordinate transformations 
have to commute. Furthermore, averaging over a region which is closely 
located around some point should, of course, reproduce the metric at this 
point.

In two respects the approach in the present paper is modest.  
Firstly we assume that a reasonable foliation into space and time has 
already been performed, so we will essentially concentrate on spatial 
averaging of a perturbed flat metric, with time kept fixed. At the end
we give a simple special generalization to four dimensions. 
Secondly
we will demonstrate covariance only in first order of the perturbation 
of the metric. 

In other respects our approach is ambitious.
 We give a closed formula for the averaging prescription, there is no need 
to solve any differential equations or to resort to a background of dust, 
perfect fluid, or whatsoever.
We not only can average over a given sphere, but may introduce an 
arbitrary smearing function $f(r)$. This appears more physical than a 
sharp cutoff at the boundary of the considered sphere. 
A very important result, at present only derived in three dimensions,
is the fact that we can apply the same formula which works for the metric
as well to the Einstein tensor. This implies that the Einstein equations for
the averaged metric are identical to the averaged equations. 

The mapping of the metric $g_{kl}(x')$ to the averaged metric $<g_{mn}>(x)$  
is 
represented by a bitensor $K_{mn}^{kl}(x'-x)$ which we will specify
in detail. A product of 
bivectors, as frequently used in the literature, is not sufficient. A mapping
with the 
help of a bitensor was also formulated by Boersma \cite{Boersma}, although 
without going into details.

The paper is organized as follows.

In sect. 2 we introduce the necessary technicalities for the projective 
coordinates. Technically these are much more convenient than the familiar 
polar coordinates, in particular for the many partial integrations which we
have to perform.
In sect. 3 we present the general form of the averaging formula. A central 
aspect is that, besides the arbitrary smearing function $f(r)$, the
integrals $F(r)$ and $G(r)$ over $f$ and $f/r$ appear. This is essential for 
the proof of covariance in sect. 4. Fixing the remaining freedom of the 
parameters appropriately, we show in sect. 5 the surprising and highly welcome
result, that our formula also yields a covariant averaging of the Einstein
tensor. In sect. 6 we discuss a simple four dimensional covariant 
generalization for 
static perturbations. The proof for the covariant averaging of the Einstein 
tensor ``almost'' goes through also in this case, but fails at the very end.
Principal limitations and possible generalizations are discussed in the 
conclusions.

\setcounter{equation}{0}\addtocounter{saveeqn}{1}%

\section{Projective coordinates}

For the moment we choose the point, where the averaging is to be performed, 
as the origin for 
simplicity. Instead of using standard 
polar coordinates $x =  r \sin \theta \cos \phi, \;
y =  r \sin \theta \sin \phi, \; z =  r \cos \theta$, it is much more 
convenient for various reasons to use  the coordinates 
$r,u\equiv u^1,v\equiv u^2$ of
stereographic projection. They are connected to the polar coordinates by
$u =  \tan \frac{\theta}{2} \, \cos \phi ,\; 
v =  \tan \frac{\theta}{2} \, \sin \phi$, 
and to the cartesian coordinates by

\begin{equation}
u\equiv u^1 = \frac{x}{r+z},\;v\equiv u^2=\frac{y}{r+z},\;r=\sqrt{x^2+y^2+z^2},
\end{equation}
or, vice versa, 

\begin{equation}
x=\frac{2u}{1+w^2} \, r,\;y=\frac{2v}{1+w^2} \, r,\;
z=\frac{1-w^2}{1+w^2} \, r, \mbox{\quad with }w^2 \equiv u^2+v^2.
\end{equation}
The geometric meaning is simple. $2\sqrt{w^2}$ is the distance from the 
north pole $(0,0,1)$  to the point where the straight line 
from the south pole $(0,0,-1)$ through $(x,y,z)$ on the unit sphere 
cuts the plane $z=1$. The 
north pole corresponds to $u=v=0$, the south pole to 
$u^2+v^2 = \infty$.
 The integration element for the angular averaging is

\begin{equation}
\frac{d\Omega}{4\pi} = \frac{1}{4\pi}\sin \theta d\theta d\phi = 
\frac{1}{\pi} \frac{dudv}{[1+w^2]^2}.
\end{equation}
The partial derivatives in projective coordinates become

\begin{equation}
\frac{\partial}{\partial x^l} =  \frac{x_l}{r} 
\frac{\partial}{\partial r} + 
\frac{\partial u^i}{\partial x^l} \frac{\partial}{\partial u^i}, 
\end{equation}
with

\begin{equation}
\frac{\partial u^i}{\partial x^l} = 
\frac{\epsilon _{3lk} \epsilon_{3ij}x_j 
- \epsilon_{ij} \epsilon_{jlk} (r+z)}{r(r+z)^2} x^k
\mbox{ (for } l=1,2),  
\frac{\partial u^i}{\partial x^3}  =  -\frac{u^i}{r}.
\end{equation}

Fortunately (2.5) will not be needed explicitely,
we will only need two simple obvious properties of the partial derivatives 
$\partial u^i/\partial x^l $: 
They are  orthogonal to $x^l$, i.e.
$x^l \; \partial u^i/\partial x^l = 0$, and proportional to $1/r$ for
fixed $u,v$.

The perturbed three dimensional metric in cartesian coordinates is written 
as $g_{mn}(x) = \delta_{mn} + h_{mn}(x)$. This decomposition is 
preserved under
 translations, rotations, and infinitesimal transformations. 
In the projective coordinates 
we also split off the flat part and write (indices $i,j$ run from 1 to 2 
in (2.6) and refer to $u^1,u^2$)

\begin{equation}
g_{rr}  =   1 + h_{rr}, \;
g_{ri}  =  \frac{2r}{1+w^2} \; h_{ri}, \;
g_{ij}  =  \frac{4r^2}{[1+w^2]^2} \; [\delta_{ij}+h_{ij}].
\end{equation}
A corresponding decomposition can be written down for the polar coordinates.

\setcounter{equation}{0}\addtocounter{saveeqn}{1}%

\section{General form of the averaging formula}

The formula presented below looks somewhat strange at first sight, therefore
it is appropriate to motivate it. Originally we tried to transform to a 
specific 
system, perform the average there, and transform back to the original one.
In this way covariance is achieved. The conditions which essentially fixed 
the specific system, denoted by primes, were 

\begin{equation}
h'_{rr}-h'_{rr}(0) = h'_{r1}-h'_{r1}(0) = h'_{r2}-h'_{r2}(0)  = 0.
\end{equation}
To avoid singularities of the transformation between the systems,
                     it was necessary to split off the
values at the origin in these conditions. One could next determine the 
transformation
leading from the original to the primed system, perform the average there,
and transform back. The fact that we had to split off the values at the 
origin had the unwanted consequence that part of the metric remained 
unaveraged. Instead of pursuing this approach further it is, however, more
useful to abstract from this original motivation and to concentrate on the
 structure which was obtained in this way. The important point is 
the following. After determining the transformation, choosing a smearing 
function $f(r)$, and performing the average, not only the original function 
$f(r)$, but also two integrals $F(r)$ and $G(r)$ appear. This will be the key
for achieving  covariance. So much for the motivation, all
the following is independent of it.

In detail, we choose a smearing function $f(r)$ with

\begin{equation}
\int _0^\infty f(r) dr = 1.
\end{equation}
Because we are in three space dimensions one should 
have $f(r) \sim r^2$ for $r \rightarrow 0$.

We will further need the function $F(r)$, the
integral over $f(r)$, as well as the function $G(r)$, the
integral over $f(r)/r$, both normalized such that they vanish at infinity:

\begin{equation}
F(r) = - \int _r^\infty f(r')dr',\;G(r) = -\int _r^\infty \frac{f(r')}{r'}dr'.
\end{equation}
Obviously one has $F(0) = -1$. 

The general form of the three dimensional averaging formula now reads

\begin{equation}
<g_{mn}>(0) 
 =  \int K_{mn}^{kl}  \;
g_{kl}(x,y,z) dr \frac{d\Omega}{4\pi},
\end{equation}
with

\begin{equation}
K_{mn}^{kl} = K^{kl}_{[f]mn}f(r)  
+  K^{kl}_{[F/r]mn}\frac{F(r)}{r} 
+  K^{kl}_{[G]mn} G(r).
\end{equation}
The structure of the tensors (tensors  in the sense of linear algebra) 
$K^{kl}_{[a]mn}, (a=f,F/r,G)$ is, 
with constant coefficients $A_{[a]}, \cdots ,F_{[a]}$,

\begin{eqnarray}
K^{kl}_{[a]mn}(x) & = & A_{[a]} 
(\delta _m^k \delta _n^l 
+\delta _n^k \delta _m^l ) 
+  B_{[a]} \delta _{mn} \delta ^{kl} 
+ C_{[a]} \delta _{mn} \frac{x^k x^l}{r^2} 
+ D_{[a]} \delta ^{kl} \frac{x_m x_n}{r^2} \nonumber\\
& &  
+ E_{[a]} [\delta_m^k \frac{x_n x^l }{r^2}
+ \delta_m^l \frac{x_n x^k }{r^2}+ 
\delta_n^k \frac{x_m x^l }{r^2}
+ \delta_n^l \frac{x_m x^k }{r^2}] 
+ F_{[a]} \frac{x_m x_n x^k x^l }{r^4}.
\end{eqnarray}
This is the most general tensor structure which is symmetric under the exchange
$m \leftrightarrow n$, and under $k \leftrightarrow l$.
Naive averaging would correspond to $A_{[f]} = 1/2$ and all the other 17 
coefficients $ B_{[f]}, \cdots ,F_{[G]}$ vanishing.

There are some restrictions for the coefficients from the beginning. 
The function $F(r)/r$ is singular at $r=0$. The expansion of $g_{kl}$ 
around $r=0$ starts with the constant $g_{kl}(0)$. For this term we can 
perform the angular averaging using

 \begin{equation}
\frac{x_m x_n}{r^2} \rightarrow \frac{1}{3}\delta_{mn}, \quad 
\frac{x_m x_n x^k x^l}{r^4} \rightarrow 
\frac{1}{15}(\delta_{mn} \delta^{kl} 
+ \delta_m^k \delta_n^l + \delta_n^k \delta_m^l)\quad\mbox{etc.}
\end{equation}
The tensor $K^{kl}_{[F/r]mn}$ has to vanish after angular 
integration  in order to avoid a singularity at $r=0$.  
Collecting the factors in front of 
 $\delta_m^k\delta_n^l+\delta_n^k\delta_m^l$,
and those in front of $\delta_{mn}
\delta^{kl}$,
this leads to the two conditions

\begin{eqnarray}
 A_{[F/r]} + \frac{2}{3} E_{[F/r]} + \frac{1}{15} F_{[F/r]} & = &0,\\
B_{[F/r]} + \frac{1}{3} C_{[F/r]} + \frac{1}{3} D_{[F/r]} 
+ \frac{1}{15} F_{[F/r]} & = & 0.
\end{eqnarray}
There are also restrictions for the coefficients which multiply $f(r)$ and 
$G(r)$. These will become relevant for proving covariance.

$K^{kl}_{[f]mn}$ is restricted by the 
transversality condition $x_l K^{kl}_{[f]mn} =0$ , which 
implies

\begin{equation}
C_{[f]}= -B_{[f]}, \; E_{[f]}= - A_{[f]}, \; 
F_{[f]} = 2 A_{[f]}-D_{[f]}. 
\end{equation}
Finally,  
$K^{kl}_{[G]mn}$ contains 
only terms $ \sim x^k x^l $, which means that 

\begin{equation}
A_{[G]} =  B_{[G]}= D_{[G]}= E_{[G]} =0.
\end{equation}
Therefore the expressions for $K^{kl}_{[f]mn}$ and $K^{kl}_{[G]mn}$ 
simplify to

\begin{eqnarray}
K^{kl}_{[f]mn} & = & A_{[f]} 
\Big[(\delta _m^k-\frac{x_m x^k}{r^2})
(\delta _n^l-\frac{x_n x^l}{r^2}) 
+ (m \leftrightarrow n)\Big] \\
& & +   (B_{[f]} \delta_{mn} 
+ D_{[f]} \frac{x_m x_n}{r^2})
(\delta ^{kl}-\frac{x^k x^l}{r^2}),\nonumber\\ 
K^{kl}_{[G]mn} & = & K_{[G]mn} \frac{x^k x^l}{r^2}, \mbox{ with } 
 K_{[G]mn}  =  C_{[G]} \delta_{mn} 
+ F_{[G]} \frac{x_m x_n}{r^2}.
\end{eqnarray}
Let us now investigate the limit of a smearing function $f(r)$ which is 
closely localized around $r=0$. The same then holds for $F(r)$ and $G(r)$. 
We may thus put $g_{kl}(x,y,z) = g_{kl}(0)$ in (3.4) and take it out in 
front of the integral. The angular averages can be performed using 
again (3.7). 
One is left with the radial integrals $\int_0^\infty f(r)dr = 1$, and
$\int_0^\infty  1 \cdot G(r) dr = - \int _0^\infty r \cdot G'(r) dr =-1$. 
The term with $F(r)/r$ does not enter, because the angular integration 
vanishes due to the conditions (3.8), (3.9) derived before.
The result has to be identical to $g_{mn}(0)$. This leads 
to two further conditions, derived from comparing the terms with $g_{mn}(0)$ 
and those with $\delta _{mn}g^j_j(0)$. Making use of the simplifications 
which arise from the restrictions for the coefficients refering to 
$f$ and $G$ one  obtains

\begin{eqnarray}
\frac{1}{15} [  14  A_{[f]} - 2 D_{[f]}-2F_{[G]}] & = & 1,\\
\frac{1}{15} [2 A_{[f]} + 10 B_{[f]} +4 D_{[f]}-5C_{[G]}-F_{[G]}]  
& = & 0.
\end{eqnarray}
The considerations above also show that the averaging of a constant 
(e.g. of $\delta _{mn}$) 
gives back this constant. This also implies, that (3.4)  can as
well be applied to the perturbation, i.e. one can replace 
$g_{mn} \rightarrow h_{mn}, \; g_{kl} \rightarrow h_{kl}$ there.

\setcounter{equation}{0}\addtocounter{saveeqn}{1}%

\section{Conditions for covariance}

Let us  apply an arbitrary infinitesimal transformation 
$x^k = x'^k + \xi ^k$, which leads to a change 
$\delta g_{kl} = \xi_{k ,l} + \xi _{l ,k} 
\rightarrow 2\xi_{k ,l}  $, when contracted with the symmetrical 
tensors in (3.6). Because  invariance with respect to translations and to 
rigid rotations around the point of consideration is manifest, one can 
restrict to transformations which leave the origin fixed, i.e. 
$\xi_k (0) = 0$, such that

\begin{equation} 
\xi _k (x) = \xi_{k,l} (0)x^l + O(r^2). 
\end{equation}
Consider the change of the integrand in (3.4). In a first step 
we transform the terms\newline 
$\sim f(r)$ and $\sim 
G(r)$  by partial integration with respect to $r$, such 
that all  three contributions become 
$\sim F(r)/r$. Because all the tensors $K^{kl}_{[a]mn}$ 
in (3.6) depend on $u,v$ only, but are independent of $r$, this is rather 
simple.

\begin{equation}
 \int _0^\infty f(r) \xi _{k,l} dr  
 =    \xi _{k,l}(0)  
- \int _0^\infty \frac{F(r)}{r} (r\frac{\partial}{\partial r} 
\xi _{k , l}) dr,
\end{equation}
where we used $F(0)=-1$ in the boundary term.

In the term with $G(r)$ we only use the factor $x^l/r$ in (3.13), 
not the $x^k/r$. 
Both of them are well defined and independent of $r$, they only depend 
on the angles (or on $u,v$, respectively) which are still fixed here.
One can apply the following chain of partial 
integrations, where we used (4.1) and $F(0)=-1$ in the 
boundary terms:

\begin{eqnarray}
\int _0^\infty G(r) \frac{x^l}{r} \xi _{k ,l}  dr 
& =  & 
\int _0^\infty G(r) (\frac{\partial}{\partial r}
\xi _k)  dr = - \int _0^\infty \frac{f(r)}{r} \xi _k  dr
\nonumber\\
& = & -\frac{x^l}{r}  \xi_{k,l}(0) 
+ \int _0^\infty \frac{F(r)}{r} (\frac{\partial}{\partial r} 
\xi _k - \frac{\xi _k}{r}) 
dr.
\end{eqnarray}
All together this leads to the integrand (to be averaged over the angles)

\begin{eqnarray}
& & 2 (K^{kl}_{[f]mn} - K_{[G]mn}
\frac{x^k x^l }{r^2})
\xi_{k,l} (0) \nonumber\\
& & + 
2 \int _0^\infty \frac{F(r)}{r}\bigg\{[-K^{kl}_{[f]mn} 
r\frac{\partial}{\partial r} 
  + K^{kl}_{[F/r]mn}] \xi _k ,_ l 
+ K_{[G]mn} \frac{x^k}{r} 
(\frac{\partial}{\partial r} \xi _k - \frac{\xi _k}{r}) \bigg\} 
 dr.
\end{eqnarray}
All manipulations which involve partial integrations with respect to $r$ 
have now been 
performed. We next  consider the curly bracket in (4.4) which has to vanish
after averaging over $u,v$. 
We introduce the projective coordinates and use

\begin{equation}
\xi _{k,l} = \bigg( \frac{x_l}{r} 
\frac{\partial}{\partial r} + 
\frac{\partial u^i}{\partial x^l} \frac{\partial}{\partial u^i} \bigg) 
\xi_k. 
\end{equation}
From (2.5) we recall that the partial derivatives $\partial u^i/\partial x^l$ 
are  orthogonal to $x^l$ and proportional to $1/r$. 
Moving $r\partial /\partial r$ to the right of 
$\partial u^i/\partial x^l$ in 
the term  $-K^{kl}_{[f]mn}(r\partial /\partial r)\xi _k ,_l $ 
thus gives an extra 
contribution.
The curly bracket in (4.4) becomes

\begin{eqnarray}
  \lefteqn{\bigg\{[ -K^{kl}_{[f]mn} (\frac{x_l}{r} 
\frac{\partial}{\partial r}
+ \frac{\partial u^i}{\partial x^l}\frac{\partial}{\partial u^i} )  
    r \frac{\partial}{\partial r} 
 +  K^{kl}_{[f]mn} 
\frac{\partial u^i}{\partial x^l} \frac{\partial}{\partial u^i}} 
\nonumber\\
 &  & + K^{kl}_{[F/r]mn} (\frac{x_l}{r} 
\frac{\partial}{\partial r}
+ \frac{\partial  u^i}{\partial x^l}\frac{\partial}{\partial u^i} )
+ K_{[G]mn} \frac{x^k}{r} 
(\frac{\partial}{\partial r}  - \frac{1}{r})
\bigg\} \xi _k .
\end{eqnarray}
The first term contains a second derivative with respect to $r$. This term
 has no chance to cancel against anything else, but it vanishes due 
to the transversality condition $x_l K^{kl}_{[f]mn} = 0$. 
We are left with terms $\sim \partial/\partial r$ and those 
$\sim 1/r$ (recall that $\partial u^i/\partial x^l \sim 1/r$). 
Both of them have to vanish after angular averaging for any $\xi_k$.
Thus we multiply by the integration element (2.3) and remove all
partial derivatives $\partial /\partial u^i$ acting on $\xi _k$ by 
partial integrations. This gives two conditions which arise from 
collecting terms  $\sim \partial/\partial r $  and terms 
$\sim  1/r$:

\begin{eqnarray}
\frac{\partial}{\partial u^i} 
\bigg(\frac{K^{kl}_{[f]mn}}{[1+w^2]^2}r 
\frac{\partial u^i}{\partial x^l} \bigg) 
+ \frac{K^{kl}_{[F/r]mn}}{[1+w^2]^2} \frac{x _l}{r} 
+ \frac{K_{[G]mn}}{[1+w^2]^2}\frac{x ^k}{r} & = & 0,
\\
-\frac{\partial}{\partial u^i} 
\bigg( \frac{K^{kl}_{[f]mn} 
+ K^{kl}_{[F/r]mn}}{[1+w^2]^2} \: r 
\frac{\partial u^i}{\partial x^l} \bigg) 
- \frac{K_{[G]mn}}{[1+w^2]^2}\frac{x ^k}{r} & = & 0.
\end{eqnarray} 
The sum gives a condition for $K^{kl}_{[F/r]mn}$ alone,

\begin{equation}
- \frac{\partial}{\partial u^i} 
\bigg(\frac{K^{kl}_{[F/r]mn}}{[1+w^2]^2}r 
\frac{\partial u^i}{\partial x^l} \bigg) +
\frac{K^{kl}_{[F/r]mn}}{[1+w^2]^2} \frac{x _l}{r} =0,
\end{equation}
therefore we will discuss (4.9) and (4.7) in the following.

One now has to use the properties of the partial derivatives 
$\partial u^i/\partial x^l$ in (2.5),
and to introduce the expressions (3.6), (3.12), (3.13) 
for the $K^{kl}_{[a]mn}$. 
The further treatment can be greatly simplified by making use of  the fact 
that, from rotation invariance and symmetry, 
the terms in (4.7) - (4.9) must be a superposition of
the form $a(r) \delta_{mn} x^k
+ b(r) (\delta_m^k x_n + \delta _n ^k x_m) + c(r) x_mx_nx^k/r^2$.
Therefore there are only three invariants which have to vanish, 
and we are free to choose simple
special cases  in order to determine them. 

The first choice is to 
 take the trace $m = n$ and put $k = 3$. This gives  
$[3a(r)+2b(r)+c(r)]z $. Next one can put 
$m = n = k = 3$ which gives $[a(r)+2b(r)+c(r)z^2/r^2]z$. The vanishing of 
these two
expressions implies already the three equations $a(r)=b(r)=c(r)=0$.
While the first choice gives only one relation, the second one gives two 
 relations from collecting the terms 
$\sim (1-w^2) /[1+w^2]^3$ and 
$\sim (1-w^2)^3/[1+w^2]^5$.
The three relations such obtained from (4.9) read

\begin{eqnarray}
6 A_{[F/r]} + 9 B_{[F/r]} + 3 C_{[F/r]} + 3 D_{[F/r]} + 4 E_{[F/r]} 
+ F_{[F/r]} 
& = & 0,\\
6 A_{[F/r]} + 3 B_{[F/r]} + C_{[F/r]} - 2 D_{[F/r]} - 2 E_{[F/r]} & = & 0,
\\
5 D_{[F/r]} + 10 E_{[F/r]} + F_{[F/r]} & =  & 0.
\end{eqnarray}
The earlier equations  (3.8), (3.9) are consequences of (4.10) - (4.12), 
therefore we can forget them.
The same procedure can be applied to (4.7) and leads to three further 
independent conditions:

\begin{eqnarray}
 4 A_{[f]} + 6 B_{[f]} + 2 D_{[f]} - 2 A_{[F/r]} - 3 B_{[F/r]} 
- 3 C_{[F/r]} - D_{[F/r]}  & & \nonumber\\
- 4 E_{[F/r]} - F_{[F/r]} 
- 3 C_{[G]} - F_{[G]} & = & 0,\\
 8 A_{[f]} + 2 B_{[f]} - 2 D_{[f]} - 2 A_{[F/r]} - B_{[F/r]} - C_{[F/r]} 
- 2 E_{[F/r]} - C_{[G]} & = & 0,\\
 8 A_{[f]} - 4 D_{[f]} + D_{[F/r]} + 2 E_{[F/r]} + F_{[F/r]} + F_{[G]} 
& =  & 0.
\end{eqnarray}
If (4.10) - (4.15) are fulfilled, the integral in (4.4), when averaged over 
the angles, respectively over $u,v$, vanishes, i.e. the expression is gauge 
invariant.

Finally we have to consider the boundary term 
$2 (K^{kl}_{[f]mn} - K_{[G]mn} x^k x^l/r^2) \xi _{k,l} (0)$ in (4.4). 
Because $\xi _{k,l} (0)$ is constant, the angular averaging 
can be performed explicitly.  In order to fulfill the covariance 
condition (1.1), the result must
 be identical to  the change of
 $<g_{mn}>(0)$ on the lhs, i.e. to $\xi _m,_n(0) + \xi _n,_m(0)$.
This results in two further conditions which are identical with (3.14), (3.15).

It is worthwhile to mention that  the covariance conditions (4.10) - (4.15)
were fulfilled in our original approach which was mentioned in the 
motivation at the
beginning of sect. 3. On the other hand, (3.14), (3.15) failed. This 
failure is due to the fact that there remained contributions 
which were not averaged. 

We have found an averaging formula which is covariant.
It contains 11 constant parameters $A_{[f]}, \cdots ,F_{[G]}$ 
(3 multiplying $f(r)$, 6 multiplying $F(r)/r$, and 2 multiplying $G(r)$),
and has to fulfill 8 independent conditions (3.14), (3.15), (4.10) - (4.15).
There is still some freedom which one can use.  For reasons to become clear 
in the next section we impose three further conditions,

\begin{equation}
C_{[a]} = D_{[a]}, \mbox{ for } a = f,F/r,G.
\end{equation}
This implies a  further symmetry of $K_{mn}^{kl}$: 

\begin{equation}
K_{mnkl} = K_{klmn}.
\end{equation}
The terms with $A,B,E,F$ obviously respect this symmetry automatically.

The  conditions now fix the  parameters uniquely. The result is\\[1ex]

\begin{tabular}{ccccccc}
 & A & B & C & D & E & F\\[0.3ex]
f&3/16&-21/16&21/16&21/16&-3/16&-15/16\\
F/r&-21/16&27/16&-63/16&-63/16&9/16&225/16\\
G&0&0&0&0&0&-15/2
\end{tabular}
\\[-10ex]

\begin{equation}
\end{equation}
\\

\setcounter{equation}{0}\addtocounter{saveeqn}{1}%

\section{Covariant averaging of the Einstein tensor}

Besides the metric, the Einstein tensor is the most important object 
 in general relativity, because it enters, together with the energy 
momentum tensor, directly  the field equations.
It would be highly desirable if one could average the Einstein tensor in 
exactly the same way as the metric tensor, and if the averaged Einstein
tensor would be identical to the Einstein tensor derived from the averaged 
metric. The old problem that the averaged equations are not identical to the 
equations with the averaged metric would then disappear.

Let us thus investigate the Einstein tensor. In first order of the 
perturbation  one has

\begin{equation}
2G_{mn}= h^i_i,_{mn}+h_{mn},^i_i - h_m^i,_{ni}- h_n^i,_{mi} 
- h^i_i,^j_j \delta_{mn} + h_{ij},^{ij}\delta_{mn}.
\end{equation}
Indices are raised and lowered with $\delta _{ij}$ here, so their position 
is in fact irrelevant.
The averaging formula (3.4) 
is now used for an arbitrary point $x$, the
integration variables are denoted by a prime, and  
$K^{kl}_{mn}(x'-x)$ depends on the difference $x'-x$. The distance $r'$ now
means $r' = |x'-x|$. 
It is convenient to introduce the modified expression 

\begin{equation}
\tilde{K}^{kl}_{mn}(x'-x) 
\equiv \frac{1}{4\pi {r'}^2} K^{kl}_{mn}(x'-x), 
\end{equation}
such that 

\begin{equation}
\int K^{kl}_{mn}(x'-x)h_{kl}(x') dr' \frac{d \Omega'}{4\pi} = 
\int \tilde{K}^{kl}_{mn}(x'-x)h_{kl}(x') d^3x'. 
\end{equation}
This makes partial integrations easy in the case under consideration. 
Differential operators 
$\partial /\partial {x'}^n$ 
acting on $h_{kl}(x')$ can be shifted
to  $\tilde{K}^{kl}_{ij}(x'-x)$ by partial integration, and finally be 
replaced by $\partial /\partial x^n$ acting on $\tilde{K}^{kl}_{ij}(x'-x)$.

There are now two possibilities of averaging:

 The first possibility is to average the metric in the way described before.   
Subsequently one calculates the Einstein tensor from (5.1), using the 
averaged metric on the rhs.
The factor of $h_{kl}$ in the integrand then becomes

\begin{equation}
I_{mn}^{kl} = \tilde{K}_i^{ikl},_{mn} + \tilde{K}_{mn}^{kl},_i^i 
- \tilde{K}_m^{ikl},_{ni} - \tilde{K}_n^{ikl},_{mi} 
- \tilde{K}_i^{ikl},_j^j\delta _{mn} 
+ \tilde{K}_{ij}^{kl},^{ij} \delta_{mn}.
\end{equation}
The second possibility is to calculate the Einstein tensor $G_{kl}$ within
the old metric and then average it with our formula in exactly the same way 
as we averaged the metric tensor. Shift the partial derivatives from the 
metric to $\tilde{K}$, and rename dummy indices where necessary such
that $h_{kl}$ appears in all six terms. The factor of $h_{kl}$ in the 
integrand  now becomes

\begin{equation}
J_{mn}^{kl} = \tilde{K}_{mn}^{ij},_{ij}\delta^{kl} 
+ \tilde{K}_{mn}^{kl},_i^i 
- \tilde{K}_{mn}^{ki},^l_i - \tilde{K}_{mn}^{li},^k_i 
- \tilde{K}_{mni}^i,^j_j \delta^{kl} + \tilde{K}_{mni}^i,^{kl}.
\end{equation}
We have to check whether the two expressions are identical. This would be an
extremely complicated task if attacked by brute force. Fortunately one can 
simplify the problem a little bit, although it stays complicated. 
An inspection of (5.4), (5.5)  shows the following property. 
The terms $\tilde{K}_{mn}^{kl},^i_i$ are identical.
We next use that the tensor
$\tilde{K}_{mnkl}$ is invariant under the exchange 
$(mn) \leftrightarrow (kl)$.
 The remaining five terms, if
arranged properly (e.g. the first in (5.4) and the last in (5.5)), 
correspond to each other by using this symmetry. This 
implies  $J_{mnkl} = I_{klmn}$. Therefore the difference 
$I_{mnkl} - J_{mnkl} = I_{mnkl} - I_{klmn}$ is antisymmetric 
under the exchange 
$(mn) \leftrightarrow (kl)$. On the other hand, the tensor composition of
this expression must 
have the general form  (3.6). The terms $\sim A,B,E,F$ are symmetric under 
the exchange $(mn) \leftrightarrow (kl)$ and thus cannot appear, therefore the 
difference has to be of the form

\begin{equation}
I_{mnkl} - J_{mnkl} = c(r') [\delta_{mn}\frac{(x'-x)_k (x'-x)_l}{{r'}^2} 
- \delta_{kl} \frac{(x'-x)_m (x'-x)_n}{{r'}^2}].
\end{equation}
The knowledge of this structure allows a considerable simplification, 
because one may now, e.g. contract $k=l$ in order to extract the function
$c(r')$. One cannot further contract $m=n$ because then the rhs
 of (5.6) vanishes identically. Instead of the four
indices present originally, one thus has to deal with two indices
only and can investigate the expression

\begin{equation}
I_{mnk}^k - J_{mnk}^k =  
c(r') [\delta_{mn} - 3 \frac{(x'-x)_m (x'-x)_n}{{r'}^2}].
\end{equation} 
It is now necessary to insert the explicit form of 
$\tilde{K}^{kl}_{mn}(x'-x)$, to 
perform the differentiations, and to make use of the relations between 
the functions $f(r),F(r)/r,G(r)$. The elementary but tedious calculation
gives the structure (5.7) with $c(r)$ a superposition of five terms which are 
proportional to $f''(r),f'(r)/r,f(r)/r^2,F(r)/r^3,G(r)/r^2$.

Let us, for the moment, keep the parameters free and only make use of the 
special forms of the tensors $K^{kl}_{[f]mn}$ and $K^{kl}_{[G]mn}$ 
 in (3.12), (3.13), as well, of course, of the conditions $C_{[a]} = D_{[a]}$
in (4.16). The result is  striking, therefore we show the explicit result here:
\begin{eqnarray}
c(r) 
& = & \quad 0 \cdot  f''(r)\nonumber\\
& & +[2 A_{[f]} + 2 B_{[f]} - B_{[F/r]} - C_{[F/r]}] \; f'(r)/r\nonumber\\
& & + [- 4 B_{[f]}  + 7 B_{[F/r]}
+ 6 C_{[F/r]} + F_{[F/r]} + F_{[G]}] \; f(r)/r^2\nonumber\\
& & + [- 15 B_{[F/r]} - 10 C_{[F/r]} - F_{[F/r]}] \; F(r)/r^3\nonumber\\
& & + 0 \cdot G(r)/r^2.
\end{eqnarray}
A sort of miracle happens. All five coefficients in front of the functions
vanish for the parameters in (4.18)! This implies that the averaged Einstein 
tensor is identical to the Einstein tensor of the averaged metric, i.e. the 
averaged field equations are identical to the field equations of the averaged
metric. 

This property is highly welcome and it is hard to believe that it is 
accidental. 
Clearly the symmetry relations shared by our 
averaging formula and by the Einstein tensor played a central role in 
the derivation of this result. This becomes evident if
one writes 

\begin{equation}
2G_{mn} = T_{mn}^{kl} h_{kl},
\end{equation}
with the operator

\begin{eqnarray}
 T_{mn}^{kl}  & = & \delta^{kl} \partial _m \partial _n 
+ \frac{1}{2} (\delta _m^k \delta _n^l + \delta _n^k \delta _m^l)  
\partial ^i \partial _i
-  \frac{1}{2} 
(\delta _m^k \partial _n \partial ^l  + \delta _m^l \partial _n \partial ^k
+\delta _n^k \partial _m \partial ^l + \delta _n^l \partial _m \partial ^k) 
\nonumber\\
& & - \delta _{mn} \delta ^{kl} \partial ^j \partial _j 
+ \delta _{mn} \partial ^k \partial ^l.
\end{eqnarray}
Both expressions, $\tilde{K}_{mnkl}$ as well as $T_{mnkl}$, are symmetric under
$m \leftrightarrow n$, under $k \leftrightarrow l$, and under 
$(m,n) \leftrightarrow (k,l)$. These symmetries implied  the vanishing of the 
five symmetric tensors in the difference of (5.4) and (5.5). The 
vanishing of the remaining antisymmetric structure in (5.6) could be 
demonstrated explicitly,
but at present we are not aware of some deeper reason behind this. 

The result for the covariant averaging of the Einstein tensor is certainly 
not trivial. For the Ricci tensor, which does not fulfill the above 
symmetry properties, the relation is not valid.

\setcounter{equation}{0}\addtocounter{saveeqn}{1}%

\section{Static perturbations in Minkowski space}

Our extension to the four dimensional case is rather modest. We  assume 
that one can find a system in which the perturbation is approximately
static. We also neglect the slow time dependence in the Robertson Walker 
metric, so one may choose coordinates such that the unperturbed metric is
the Minkowski one, (-1,1,1,1). To keep this situation, only rigid 
translations, 
rigid spatial rotations, and infinitesimal transformations which keep the 
time unchanged are allowed. This means that $\xi^0 = 0$, and $\xi ^m$ is
independent of $t$. Furthermore we can drop all time derivatives in the metric.
Under these restrictions the perturbations $h_{00}$ and $h_{m0}$ become 
gauge invariant.

We average the perturbation with the following simple ansatz. 

\begin{eqnarray}
<h_{mn}>(0) & = & \int  \tilde{K}_{mn}^{kl}  \; h_{kl} 
 r^2 dr \frac{d\Omega}{4\pi}, \\
<h_{00}>(0) & = & \int p(r)
h_{00}  r^2 dr \frac{d\Omega}{4\pi},\\
<h_{m0}>(0) & = & \int q(r)
h_{m0}  r^2 dr \frac{d\Omega}{4\pi}.
\end{eqnarray}
Of course the functions $p(r)$ and $q(r)$ have to fulfill the 
normalization conditions 
\begin{equation}
\int _0^\infty p(r) r^2 dr  = \int _0^\infty q(r) r^2 dr = 1.
\end{equation}
The equations (6.1) - (6.3) are written in such a way that the volume element
$d^3x/4\pi$ appears in the integrals. 
Hopefully, the fact that we sometimes include the factor $r^2$ of the
volume element into the 
function (as in $f(r)$), and sometimes don't (as in $p(r),q(r)$), does not
produce too much confusion. The choice is motivated by the way how we have 
to perform the various partial integrations.

Equations (6.1) - (6.3) are a simple generalization 
of our previous formula. For static perturbations they are covariant
in the sense of (1.1) with respect 
to static transformations. One could use a more general ansatz, where a term
with $h^0_0$ is inserted into the rhs of the averaging formula (6.1),
and a 
gauge invariant combination of $h^k_k$  and $(x^kx^l/r^2) h_{kl}$
 into (6.2). We found that this does not help to solve the problem 
which will arise at the end, therefore we keep things simple and work 
with (6.1) - (6.3).

Let us now consider the (lowest order) Einstein tensor which reads

\begin{eqnarray}
2 G_{mn} & = & 2 G_{mn}^{(s)} + h^0_0,_{mn} -  h^0_0,^i_i\delta_{mn} ,\\
2 G_{00} & = & h^i_i,^j_j - h_{ij},^{ij} \\
2 G_{m0} & = & h_{m0},^i_i - h_{i0},_m^i.
\end{eqnarray}
Here $G_{mn}^{(s)}$ is the spatial part of the Einstein tensor in (5.1). 
Under the assumptions above, the additional terms in $G_{mn}$,  as well as
 $G_{00}$ and $G_{m0}$ 
are invariant under infinitesimal static transformations.

We now investigate whether  the Einstein 
tensor of the averaged metric can be identical to the averaged Einstein tensor.
For $G_{m0}$ this is trivial,  
we start with $G_{mn}$. For the  part $G_{mn}^{(s)}$ 
 we know from the previous 
section that the result is independent of the order of averaging. We can
restrict to the additional contributions in (6.5).

If we average the perturbation $h_0^0$ 
according to (6.2) and introduce into (6.5)  we obtain the integrand

\begin{equation}
 [p,_{mn} - p,_i^i \delta_{mn}]h^0_0 = [(p'' - \frac{p'}{r})\frac{x_mx_n}{r^2} 
- (p'' + \frac{p'}{r})\delta_{mn}]h^0_0.  
\end{equation}
If, alternatively, we first calculate $2 G_{mn}$ in (6.5) with the old metric 
and then average it in the same way as (6.1), i.e. replace 
$h_{kl}$ by $G_{kl}$  there on the rhs, and
shift the partial derivatives from the perturbation to the multiplying 
functions, we obtain the integrand

\begin{equation}
[\tilde{K}_{mn}^{kl},_{kl} - \tilde{K}_{mnk}^k,_l^l]h^0_0.
\end{equation}
This gives the condition

\begin{equation}
\tilde{K}_{mn}^{kl},_{kl} - \tilde{K}_{mnk}^k,_l^l  =  
  (p'' - \frac{p'}{r})\frac{x_mx_n}{r^2} 
- (p'' + \frac{p'}{r})\delta_{mn}.
\end{equation}
We next apply the same procedure to $G_{00}$. If we average the 
perturbations $h_i^i$ and $h_{ij}$ 
according to (6.1) and introduce into (6.6)  we obtain the integrand

\begin{equation}
[\tilde{K}_i^{ikl},_j^j - \tilde{K}_{ij}^{kl},^{ij}]h_{kl}.
\end{equation}
If, alternatively, we first calculate $2 G_{00}$ in (6.6) with the old metric 
and then average 
it in the same way as (6.2), i.e. replace $h_{00}$ by $G_{00}$ there,
and shift the partial derivatives from the perturbation to the multiplying 
functions, we obtain the integrand

\begin{equation}
[ p,_j^j \delta^{kl} - p,^{kl} ]h_{kl} 
= - [(p'' - \frac{p'}{r})\frac{x^kx^l}{r^2} 
- (p'' + \frac{p'}{r})\delta^{kl}]h_{kl}.
\end{equation}
The condition $\tilde{K}_i^{ikl},_j^j - \tilde{K}_{ij}^{kl},^{ij} =
- [(p'' - \frac{p'}{r})\frac{x^kx^l}{r^2} 
- (p'' + \frac{p'}{r})\delta^{kl}]$ which arises from  (6.11), (6.12) 
is a consequence 
of  (6.10) if  one renames the dummy indices $k,l$ there, subsequently
replaces $m,n$ by $k,l$, and uses
the symmetry $\tilde{K}_{mnkl} = \tilde{K}_{klmn}$. We thus only need
to consider (6.10) in the following.

With (3.5), (3.6) and the parameters in  (4.18), a lengthy but 
elementary calculation gives

\begin{equation}
\tilde{K}_{mn}^{kl},_{kl} - \tilde{K}_{mnk}^k,_l^l = K_1(r) 
\frac{x_mx_n}{r^2} + K_2(r) \delta_{mn},
\end{equation}
with 

\begin{eqnarray}
K_1(r) & = & -\frac{9}{4} \frac{f''(r)}{r^2} 
 + \frac{15}{2} \frac{f'(r)}{r^3}
+ \frac{9}{2} \frac{f(r)}{r^4} 
 - 30  \frac{G(r)}{r^4},\\
K_2(r) & = & \frac{9}{4} \frac{f''(r)}{r^2} 
 - 3 \frac{f'(r)}{r^3}
 - 6 \frac{f(r)}{r^4} 
+ 15 \frac{G(r)}{r^4}.
\end{eqnarray}
If one equates the factors
of $x_mx_n/r^2$ and $\delta_{mn}$ in  (6.10) one obtains the two conditions

\begin{eqnarray}
K_1(r) & = & p''(r) - p'(r)/r,\\
K_2(r) & = & - p''(r) - p'(r)/r.
\end{eqnarray}
The sum of these equations gives

\begin{equation}
p'(r) = - \frac{r}{2} (K_1(r) + K_2(r)).
\end{equation}
If one calculates $p''(r)$ from this, and inserts  again into
(6.16), (6.17) one obtains the integrability condition

\begin{equation}
K_1'(r) + K_2'(r) + \frac{2 K_1(r)}{r} = 0.
\end{equation}
This condition consists, in fact, of six conditions, i.e. all the constants
in front of $f'''(r)/r^2,f''(r)/r^3,f'(r)/r^4,f(r)/r^5,F(r)/r^6,G(r)/r^5$
have to vanish. Again a sort of miracle occurs. All these constants vanish
for the parameters in (4.18).

Up to now everything worked perfectly,  but now we run in trouble with 
the normalization condition (6.4). From (6.18) one gets 
 
\begin{eqnarray}
1 & \stackrel{!}{=} & \int _0^\infty r^2 p(r) dr 
=  \Big[\frac{r^3}{3} p(r) \Big]_0^\infty 
- \frac{1}{3} \int_0^\infty r^3 p'(r)dr \nonumber\\
& = & 
\frac{1}{6} \int _0^\infty r^4 (K_1(r) + K_2(r))dr =  \frac{3}{2}.
\end{eqnarray}
Unfortunately the normalization
becomes $3/2$ and not 1. 
The boundary values $[r^3 p(r)]_0^\infty$ 
cannot help, because these have to vanish in order that 
$\int_0^\infty r^2p(r)dr$ 
converges.

 Thus we ``almost''
succeeded  to extend the applicability of the covariant averaging formulae 
(6.1) - (6.3) to the Einstein tensor, but we failed
at the end. There is a clash between the condition from covariance and 
the normalization. A slightly more general approach can probably solve 
this problem.

\setcounter{equation}{0}\addtocounter{saveeqn}{1}%

\section{Outlook and conclusions}

The covariant averaging procedure presented in this paper is 
complicated. This was to be anticipated, one could not expect to obtain 
a simple solution to a complicated problem. 

Compared to most approaches in the literature our formula shows the 
following features. First one may be surprised that there is no 
factor $\sqrt{g}$ in the integrand. We have no comment on this, 
it simply is not present. 
An important point is, that it is not sufficient to work with bivectors 
which mix the indices.
It is necessary to have bitensors in (3.6). Our formula is more general
than prescriptions which only average over a certain volume because we can
use an arbitrary normalized smearing function $f(r)$. But it is essential 
that, besides the function $f(r)$, also the integrals $F(r)$ and $G(r)$ 
appear. Two simple examples 
suggest themselves. We recall that $f(r)$ should behave $\sim r^2$ for 
small $r$. \newline
Averaging over a sphere of radius $r_0$:

\begin{equation}
 f(r) = \frac{3 r^2}{r_0^3}\Theta(r_0-r), \; 
F(r) = (-1 +\frac{r^3}{r_0^3})\Theta(r_0-r) , \;
G(r) = (- \frac{3}{2r_0} + \frac{3 r^2}{2r_0^3})\Theta(r_0-r).
\end{equation} 
Averaging with an exponentially decreasing function:
\begin{equation}
 f(r) = \frac{ r^2}{2r_0^3} e^{-r/r_0}, \; 
F(r) = - (1 + \frac{r}{r_0} + \frac{r^2}{2r_0^2})e^{-r/r_0} , \;
G(r) = - ( \frac{1}{2r_0} + \frac{ r}{2r_0^2}) e^{-r/r_0}.
\end{equation} 
One has to face the fact that there is  no chance to find a much simpler 
covariant averaging formula in three dimensions than the one presented here.
If one works in first order of the perturbed metric, there must be a linear 
connection between the original perturbation $h_{kl}(x)$ and 
the averaged $<h_{mn}>(0)$, represented
by a tensor (tensor in the sense of linear algebra) $K_{mn}^{kl}$ which 
is symmetric with respect to 
$m \leftrightarrow n$ and to $k \leftrightarrow l$. Such a connection has also 
been discussed by Boersma \cite{Boersma}. The only 
objects which are available for
the construction of $K_{mn}^{kl}$  are the vector $x$ and Kronecker deltas. 
Therefore one ends up with the six tensors defined in (3.6). The functions 
in front of the tensors must depend on $r$ only. They are related by 
covariance. A special solution has been presented in this paper. 
There are also   more general solutions not mentioned here, e.g. averaging 
formulae which,
 besides $f(r),F(r)/r,G(r)$, also
contain the derivative $rf'(r)$. These  may
allow to find a prescription which is also suited for a 
covariant averaging of the Einstein tensor in four dimensions (for static 
perturbations). The most general form of a covariant averaging formula is 
under investigation.

A feature of our averaging formula, which might be considered as unpleasant, 
is 
the appearance of the integrals $F(r)/r$ and $G(r)$, with $F(0) = -1$ and 
$G(0)$ finite. Although the term with $F(r)/r$ does not cause a singularity
at the origin because the angular integration vanishes there, it somehow 
hampers the smoothing procedure. Before trying to find a
``better'' solution one should, however, take notice of the following
 fact. It is inevitable that 
  functions  appear  which are not too smooth at the origin. This 
should be
clear from our proof of covariance in sect. 4. An infinitesimal 
transformation within the averaging formula must result in the corresponding
transformation of the averaged metric at the origin. Technically this can 
only arise through boundary terms at zero which originate from partial 
integrations. Something substantial must be present  near zero in order to
produce these boundary terms.

An obvious task to be done is to investigate iterations of our averaging 
formula and to check whether
the iteration procedure converges. But it is clear that one will not 
always obtain a smooth metric in the limit. The reason is
again covariance. We are still free to perform gauge transformations, and by an
unfavorable choice of gauge the ``smoothed'' metric can look wavy and 
irregular. All one can expect is that the final metric becomes equivalent 
to a smooth metric.

One could also proceed to extend the approach
 to second order in the perturbation. 
We recall that (3.4) also holds if $g_{mn}$ and $g_{kl}$ are replaced by
$h_{mn}$ and $h_{kl}$. In second order an additional contribution 
which is quadratic in the perturbation will probably be needed in the integral.
Though certainly tedious, such an extension appears feasible.

A particularly pleasant property of our formula is the fact that it likewise 
yields 
 a covariant averaging of the Einstein tensor and thus of the energy momentum 
tensor in three dimensions, an extra bonus which, quite surprisingly, came out 
from the suggested averaging formula. We hope to resolve the minor 
problems found in the four dimensional case by a slightly more general 
prescription.

{\bf Acknowledgement: } I thank Juliane Behrend for valuable discussions 
and for her interest in this work.

\end{document}